\documentclass[aps,twocolumn,prb]{revtex4}

\usepackage{graphicx}
\usepackage{amssymb}       
\usepackage{amsmath}

\newcommand{\R}{\mathbf{r}}

\begin{document}

\title{Gradient-dependent upper bound for the exchange-correlation energy and application to density 
functional theory}
\author{Lucian A. Constantin}
\affiliation{Center for Biomolecular Nanotechnologies @UNILE, Istituto Italiano di Tecnologia, Via 
Barsanti, I-73010 Arnesano, Italy}   
\author{Aleksandrs Terentjevs}
\affiliation{Center for Biomolecular Nanotechnologies @UNILE, Istituto Italiano di Tecnologia, Via 
Barsanti, I-73010 Arnesano, Italy}
\author{Fabio Della Sala}
\affiliation{Istituto  Nanoscienze-CNR, Via per Arnesano 
16, I-73100 Lecce, Italy }
\affiliation{Center for Biomolecular Nanotechnologies @UNILE, Istituto Italiano di Tecnologia, Via 
Barsanti, I-73010 Arnesano, Italy}
\author{Eduardo Fabiano}
\affiliation{Istituto  Nanoscienze-CNR, Via per Arnesano 
16, I-73100 Lecce, Italy }
\affiliation{Center for Biomolecular Nanotechnologies @UNILE, Istituto Italiano di Tecnologia, Via 
Barsanti, I-73010 Arnesano, Italy}
\email{eduardo.fabiano@nano.cnr.it}

\begin{abstract}
We propose a simple gradient-dependent bound for the exchange-correlation energy (sLL),
based on the recent non-local bound derived by Lewin and Lieb. 
We show that sLL is equivalent to the original Lieb-Oxford bound
in rapidly-varying density cases but it is tighter for slowly-varying density
systems. To show the utility of the sLL bound we apply it to 
the construction of simple
semilocal and non-local exchange and correlation functionals. In both
cases improved results, with respect to the use of Lieb-Oxford bound, are obtained showing the power of the sLL bound.
\end{abstract}

\maketitle

The Lieb-Oxford (LO) bound \cite{LO1},
$E_{xc}[n_\lambda]\geq LO[n_\lambda]$ with $n_\lambda(\R)=\lambda^3n(\lambda \R)$
being the uniformly scaled density and
\begin{equation}
LO[n_\lambda] = -1.68\int n_\lambda^{4/3}(\R)d\R\ ,
\end{equation}
is one of the main exact constraints for the exchange-correlation (XC)
energy and is used in the construction of 
many semilocal XC approximations 
\cite{pbe,pbesol,apbe,pbeint,wc,tpss,revtpss,bloc,blochole,MdelCampo2012179}.
Several works have concerned the tightening of this bound. 
Chan and Handy \cite{CH} proposed to reduce the constant in LO to
1.6358. Odashima and Capelle \cite{LO2}
used numerical analysis 
to further reduce it to 1.444. 
Conversely, considering the low-density limit
of the spin-unpolarized uniform electron gas,
it is found that the constant cannot be smaller
than 1.437 \cite{P1}.
Consequently the value 1.4 has been employed to build some
recent XC functionals \cite{sogga,mggams,hyper1}.

Very recently Lewin and Lieb \cite{LL1} derived a different bound,
having a density dependence, i.e. $E_{xc}[n_\lambda]\geq LL[n_\lambda]$ where
\begin{eqnarray}
LL[n_\lambda] & = & -1.451\int n_\lambda^{4/3}(\R)d\R - \\
\nonumber
&& -0.327\left[\int\left|\nabla n_\lambda\right|d\R\right]^{1/4}\left[\int n_\lambda^{4/3} 
d\R\right]^{3/4}\ .
\end{eqnarray}
Note that both the LO and LL bounds scale as the exchange
energy under a uniform scaling of the density.
However, the LL bound, in the present form, is
not very practical for development in density functional theory.
Thus, in this communication we consider a simplified form for 
such a bound which, at the same time, tightens it, such
as to provide a useful tool for XC functional development.

To start we note that the LL bound is tighter than the LO one
only for slowly-varying densities, where $|\nabla n_\lambda|$ 
is small. Let first consider the XC energy in this regime (spin-unpolarized case), in the 
low-density limit ($\lambda\rightarrow0$), where the second-order 
gradient expansion of the XC energy \cite{HL1} becomes exact 
\begin{eqnarray}
\nonumber
 E_{xc}[n_\lambda] &=&  -1.437\int n_\lambda^{4/3} d\R - 0.091\int n_\lambda^{4/3}s^2 d\R \\
&+&  \int n_\lambda\beta t_\lambda^2 d\R \geq\\
\nonumber
& \geq & -1.437\int n_\lambda^{4/3} d\R - 0.091\int n_\lambda^{4/3}s^2 d\R \geq\\
\label{e3}
& \geq & -1.451\int n_\lambda^{4/3} d\R - 0.091\int n_\lambda^{4/3}s^\alpha d\R\ ,
\end{eqnarray}
for any $\alpha\leq 2$, and 
where $s=|\nabla n_\lambda|/[2(3\pi^2)^{1/3} n_\lambda^{4/3}]$ and
$t_\lambda=|\nabla n_\lambda|/[2(3/\pi)^{1/6}n_\lambda^{7/6}]$ are the reduced gradients for
exchange and correlation respectively
(note that $s$ is invariant under the uniform scaling while
$t_\lambda\propto\lambda^{1/2}$) and
$\beta\geq0$ is the second order expansion coefficient for correlation.
Note that $\beta$ is known only in the high-density limit, and 
in the metallic range \cite{HL1} (i.e. $2\leq r_s\leq 6$, with $r_s=[3/(4\pi n)]^{1/3}$ being the 
bulk parameter).
The last inequality uses the fact that $s\leq1$ in the considered limit.
Eq. (\ref{e3}) suggests that in the low-density limit 
a gradient-dependent term may play an important 
role to the bound, similar with the LL expression.  

On the other hand, starting form the LL bound and using the H\"older
inequality we find
\begin{equation}\label{e4}
E_{xc}[n_\lambda]\geq -1.451\int n_\lambda^{4/3} d\R - 0.516\int n_\lambda^{4/3}s^{1/4} d\R\ .
\end{equation}
This provides a slightly tighter bound than LL (not mathematically 
rigorous but in the spirit of Ref. \onlinecite{LO2}, as explained below), 
and it can be used
together with Eq. (\ref{e3}) to obtain a more useful formula for the bound. 

In fact, a comparison of Eqs. (\ref{e3}) and (\ref{e4}) suggests that the correct
bound could be
\begin{equation}\label{e5}
sLL[n_\lambda] = -1.451\int n_\lambda^{4/3} d\R - C\int n_\lambda^{4/3}s^{1/4} d\R\ ,
\end{equation}
with $0.09\leq C\leq0.52$ (to be fixed later).
A similar idea is also proposed in Ref. \onlinecite{burke}.
This ansatz is also supported by a further analysis of the
XC energy under the uniform scaling to the low-density limit
at any value of $s$ (note that instead Eq. (\ref{e3}) holds only
for $s<1$). To see this, consider the partitioning the XC energy as
$E_{xc}[n_\lambda] = E_{xc}^{LDA}[n_\lambda] + E_{xc}^{GE2}[n_\lambda] + \delta E_x[n_\lambda] + 
\delta E_c[n_\lambda]$,
where $LDA$ denotes the local density approximation,
$GE2$ denotes the second-order gradient expansion terms,
and $\delta E_x$ and $\delta E_c$ indicate higher-order
exchange and correlation contributions.
Because in this limit $t\rightarrow0$ we can set $\delta E_c\approx0$.
Moreover, following Ref. \onlinecite{revtpss}, we can assume that
$E_{xc}^{GE2}[n_\lambda]\approx0$ for $\lambda\rightarrow 0$. Hence,
\begin{eqnarray}\label{e6}
E_{xc} & \approx & E_{xc}^{LDA}[n_\lambda] +  \delta E_x[n_\lambda] =  \\
\nonumber
& = & -1.44\int n_\lambda^{4/3} d\R - C_X\int_{s\geq 1} n_\lambda^{4/3}[f_x^{exact}(s)-1] d\R\ ,
\end{eqnarray}
with $C_x=(3/4)(3/\pi)^{1/3}$ and $f_x^{exact}$ is the exact exchange enhancement factor.
Noting that at large gradients $f_x^{exact}-1 \geq 0$, Eq. (\ref{e6}) shows that a tighter 
LO bound may fail in the low-density regime 
(where both LDA correlation and exchange scale similarly, 
as $\sim 1/r_s$) and an $s$-dependent term is instead required.

To fix the parameter $C$, we consider a tough case for
the semilocal sLL bound: the beryllium isoelectronic
series  up to 
the nuclear charge $Z=140$.
This is in fact a rapidly-varying
high-density limit case with strong correlation
($E_c^{LDA}\ge E_c^{exact}\rightarrow-\infty$ for highly charged
ions \cite{Be1}).
While the original LO bound gives $LO/E_{xc}\approx 1.9$, the 
LL bound is much looser, having $LL/E_{xc}\approx 2.3$ 
(note that all results are practically independent on
the nuclear charge).
Hence, our constraint is that sLL$\approx$ LO
for such a case (that is representative for rapidly-varying density regime).
In this way we obtain $C=0.245$. 

The bound of Eq. (\ref{e5}) can be tested for
different systems. In case of atoms, ions,
molecules, and solids this test is trivial
because sLL is slightly looser than the local
bound of Odashima and Capelle \cite{LO2}
which was
verified for these cases.
Hence, we consider here a few additional possibilities,
based on interesting model systems.

\begin{figure}
\includegraphics[width=\columnwidth]{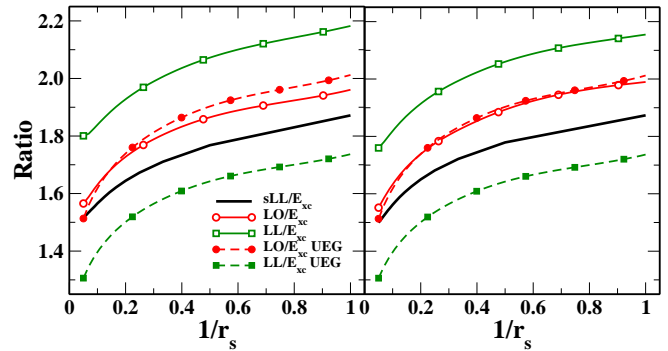}
\caption{\label{fig1} The ratios $LO/E_{xc}$, $LL/E_{xc}$, and $sLL/E_{xc}$ versus $1/r_s$, for 
40$e^{-}$ jellium spheres (left panel) and for two interacting jellium slabs (right panel). Also 
shown are $LO/E_{xc}$ and $LL/E_{xc}=sLL/E_{xc}$, for the uniform electron gas (UEG).}
\end{figure}

For slowly-varying density cases the sLL
bound is surely verified in the low-density
limit, because $C=0.245\geq0.091$.
For other density regimes we consider
in Fig. \ref{fig1}
jellium spheres with
40 electrons and bulk parameter
included in the interval $1\leq r_s\leq20$, as well as
two interacting jellium slabs of thickness
and separation distance equal to the Fermi wave length
(similar results are obtained also for other thicknesses and
distances) and $1\leq r_s\leq 20$.
The latter describe several density regimes, ranging
from slowly- (for thick and close slabs) to very rapidly-varying
ones (thin slabs).
%
%

In all cases the exchange energy is computed exactly,
whereas the correlation energy is estimated using the
JS functional \cite{js}
(similar results are given also by PBE \cite{pbe} 
and TPSS \cite{tpss}). 
The LO, sLL, and LL have been calculated using exact exchange densities.
These systems are relevant to understand the significance 
on the bound of quantum
oscillations in ordinary matter.
Figure \ref{fig1} shows that in all cases
the sLL bound is respected ($sLL/E_{xc}\geq 1$)
and it is tighter than LO. 
Instead, the non-local LL bound is always rather loose.
Finally, comparison with uniform electron gas
results shows that quantum oscillations and surface effects
have a negligible effect on the LO bound, which does not depend on the
gradient, while they give a little weakening
of the semilocal sLL bound. 

As a different example we consider the one-electron densities
\begin{equation}
n_H(r) = \frac{e^{-2r}}{\pi}\quad ;\quad n_G(r)=\frac{e^{-r^2}}{\pi^{3/2}}\ .
\end{equation}
These are models for rapidly-varying density systems and have been
shown to be relevant for bonding properties of molecules
\cite{zpbeint}. 
The calculations are analytical and we 
readily find:
$LO/E_{XC}=1.55$, $LL/E_{XC}=1.82$, and $sLL/E_{xc}=1.56$ for H;
$LO/E_{XC}=1.54$, $LL/E_{XC}=1.80$, and $sLL/E_{xc}=1.54$ for G.
Thus, the LO and sLL bounds have the same quality, both outperforming the
LL bound. The coincidence of LO and sLL for these model
systems is reminiscent of the fact that, to fix C in Eq. (\ref{e5}),
we constrained sLL to reproduce LO for the Be series, which
presents rapidly-varying densities. However, notably in the
present case there is no contribution from correlation.

To conclude our analysis we consider finally the opposite case
of slowly-varying strong-correlated systems.
One example of such systems is given by the 
point-charge-plus-continuum (PC) model \cite{pc1,pc2}. 
At small reduced gradients $s\leq 1$, the XC energy is
\cite{pc2,isi1,isi3}
\begin{equation}
E_{xc}^{PC}[n]=-1.451 \int n^{4/3}d\R + 0.203 \int n^{4/3}s^2d\R\ .
\end{equation}
We observe that the local term is exactly the one
entering in the LL and sLL bounds and is very close
to the numerical estimation of Odashima and Capelle \cite{LO2},
although this latter bound is formally broken for $s\rightarrow0$.
This shows that this model is the true limit for constant densities.
On the other hand, because the second-order XC contribution is
positive, all the examined bounds are valid for the PC model
for $0<s\leq1$.

Another interesting example of slowly-varying
strongly correlated system can be
obtained by considering the strong-correlation
scaling of the density \cite{scaling}, 
$n_\lambda(\R) = \lambda^{-2}n(\lambda^{-1}\R)$, 
in the limit $\lambda\rightarrow\infty$.
Under this scaling the reduced gradients
behave as $s_\lambda(\R) = \lambda^{-1/3}s(\lambda^{-1}\R)$
and $t_\lambda(\R)=\lambda^{-2/3}t(\lambda^{-1}\R)$.
The XC energy is dominated by the local terms,
while the second-order contributions decay as $\lambda^{-1/3}$. 
On the other hand, we have
$sLL = -1.451\lambda^{1/3}\int n^{4/3}d\R - 0.245\lambda^{1/4}\int n^{4/3}s^{1/4}d\R$.
Therefore, for $\lambda\rightarrow\infty$
sLL provides a good bound to the XC energy 
(note that the local part of sLL is almost the exact limit
for the local XC energy).
Moreover, we note that in this limit $sLL\leq LO$, being
a tighter bound for the XC energy of slowly-varying strongly-correlated
systems.

To show examples of the practical 
utility of the sLL bound
we consider its application in the construction of
a GGA and an hyper-GGA functional.
\par
\begin{figure}
\includegraphics[width=\columnwidth]{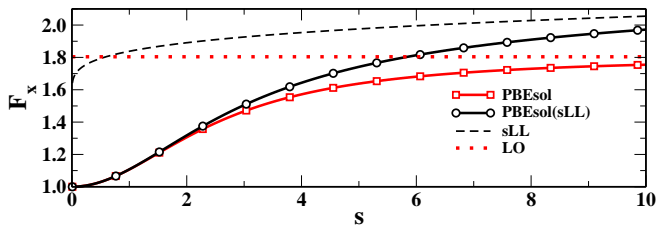}
\caption{Exchange enhancement factors versus $s$, for PBEsol and PBEsol(sLL)
Also shown, are the asymptotes of sLL and LO bounds.
}
\label{f2}
\end{figure}

 
\textbf{GGA}. We consider the new sLL bound in the PBEsol \cite{pbesol} XC
functional, which is a non-empirical GGA with the correct second-order
exchange coefficient.
The value of $\kappa=0.804$ in the exchange enhancement factor 
$F_x=1+\kappa-\kappa/[1+(\mu/\kappa)s^2]$
is replaced by
\begin{equation}
\kappa = 0.559 + 0.279s^{1/4}\ . \label{ref:eqdd}
\end{equation}
in order to  respect the 
sLL bound at any point of the space. 
For the correlation part we use $\beta=0.045$ (note that in PBEsol $\beta=0.046$) 
refitting jellium surfaces.\cite{pbesol}.
In Fig. \ref{f2}, we show a comparison of the exchange enhancement factors. 
Note that any exchange functional that recovers the local LO bound (e.g. PBEsol, TPSS meta-GGA 
\cite{tpss}), satisfies also the local sLL bound. Instead, 
PBEsol with the sLL bound clearly violates the local LO bound. 
%
%

To test the effect of the sLL bound 
we performed different tests for solids and molecules.
In Table \ref{tab1} we report the mean absolute errors (MAEs)
for solid-state tests on 24 bulk solids (see computational details).
We see that using the sLL bound  improves the results: 
lattice constants  and cohesive energies are improved with respect PBEsol, 
while bulk moduli are almost comparable.
We recall that PBEsol is one of the most used semilocal 
functionals in solid-state physics and, one of the best for lattice constants 
and bulk moduli \cite{sol1}.
AM05\cite{AM05} performs slightly worse than PBEsol \cite{sol1}
(MAEs on this test sets are 32 m\AA{} and 8.6 Gpa, respectively).
A PBEsol functional with $\kappa=0.559$  is definitely 
worse \cite{haasfamily,mukappa}, having MAEs of 35 m\AA{}, 8.1 GPa and 0.51
eV/atom. 
This shows the importance of the non-local term of the sLL bound.

\begin{table}
\begin{center}
\caption{\label{tab1} Mean absolute error (MAE) for the 
lattice constants (m\AA), bulk moduli (GPa), and cohesive energies (ev/atom) as obtained 
from the PBEsol functional with the original LO bound, and with the new sLL bound, for various types of solids. 
Reference data from Ref. \onlinecite{sol1}. 
}
\begin{ruledtabular}
\begin{tabular}{lrr}
 & \multicolumn{2}{c}{PBEsol}\\
\cline{2-3}
                   &  LO        & sLL \\  
\hline
\multicolumn{3}{c}{Lattice constants (m\AA)} \\
simple metals     & 37.6 & 31.8 \\
transition metals &  24.9 & 23.2 \\
semiconductors    &  29.6 & 31.8 \\
ionic solids      & 20.4 & 22.2 \\
insulators        &  6.5 & 6.3 \\
total MAE         &  24.5 & 23.8 \\
\multicolumn{3}{c}{bulk moduli (GPa)} \\
simple metals     & 1.0 & 0.9 \\
transition metals &  18.7 & 18.5 \\
semiconductors    &  7.8 & 8.1 \\
ionic solids      &  4.1 & 5.0 \\
insulators        &  6.7 & 6.7 \\
total MAE         & 7.7 & 7.9 \\
\multicolumn{3}{c}{cohesive energies (eV)} \\
simple metals     &  0.15 & 0.13 \\
transition metals &  0.71 & 0.64 \\
semiconductors    &  0.31 & 0.26 \\
ionic solids      & 0.10 & 0.06 \\
insulators        & 0.65 & 0.60 \\
total MAE         & 0.37 & 0.33 \\
\end{tabular}
\end{ruledtabular}
\end{center}
\end{table}

In Table \ref{tab2} we report
the MAEs for several tests on molecular properties
that are relevant for semilocal functionals.
\begin{table}
\begin{center}
\caption{\label{tab2} Mean absolute errors (MAE) in kcal/mol for energy tests, and in m\AA{} for 
geometry tests, for several molecules benchmark, as obtained 
from the PBEsol functional with the original LO bound, and with the new sLL bound.}
\begin{ruledtabular}
\begin{tabular}{lrr}
 & \multicolumn{2}{c}{PBEsol}\\
\cline{2-3}
test     &   LO     &   sLL       \\ 
\hline
atomiz. energy (AE6)    & 34.5 & 32.1 \\
atomiz. energy (W4)     & 21.4 & 20.0 \\
atomiz. energy (W4-MR)  & 35.5 & 33.9 \\
3$d$-metals AE (TM10AE) & 18.3 & 17.6 \\
Au clusters AE (AUnAE)  & 3.6 & 3.1 \\
reaction ener. (OMRE)   & 8.1 & 10.7 \\
reaction ener. (BH76RC) & 6.3 & 6.0 \\
barrier heights (BH76)  & 12.2 & 11.8 \\
3$d$-metals RE (TMRE)   & 9.9 & 9.1 \\
ionization pot. (IP13)  & 2.3 & 2.1 \\
proton aff. (PA12)      & 1.5 & 1.5 \\
difficult cases (DC9/12) & 82.9 & 78.3 \\
isomerization ener. (ISOL6) & 1.7 &  1.7 \\
interfaces (SI12)        & 3.8 & 3.5 \\
hydrogen bonds (HB6)     & 1.6 & 1.2 \\
dihydrogen bonds (DHB23) & 1.8 & 1.5 \\
dipole-dipole (DI6)      & 1.0 & 0.7 \\
bond lengths (MGBL19)    & 10.0 & 9.9 \\
total energy (AE17)      & 250.3 & 246.6 \\ 
\hline
$\sum_i {\rm MAE_i}/{\rm MAE_i^{PBEsol}}$                   & 1.0 & 0.94 \\
\end{tabular}
\end{ruledtabular}
\end{center}
\end{table}
In this case PBEsol with the new sLL bound is better than PBEsol for all tests 
(the only exception is the  OMRE test) with an overall 
relative MAE of 0.94.
The improvement traces back to the larger non-locality
of the sLL bound with respect to the LO one, which provides a better description
when large reduced gradients are involved \cite{mukappa,htbs,revpbe}.
Note that a further improvement for molecular
properties can be obtained by replacing the PBEsol correlation
with the zPBEsol one \cite{zpbeint,zvpbeint}, which is not modifying the
sLL bound behavior but improves most of the energetic properties
\cite{ourarxiv}.

\par
\textbf{Hyper-GGA.}  We consider the
hyper-GGA correlation functional \cite{hyper1}
\begin{equation}
E_c[n]=\int \frac{e_c^{TPSS}(\R)}{X(\R)-e_x^{TPSS}(\R)}(X(\R)-e_x(\R))d\R\ ,
\label{e19}
\end{equation}
where $e_c^{TPSS}$ and $e_x^{TPSS}$ are the TPSS 
correlation and exchange energy densities \cite{tpss} respectively, 
$e_x(\R)$ is the exact exchange energy density 
(we use only the conventional gauge \cite{hyper1}), and $X(\R)$ is a 
local upper bound for the $e_x^{TPSS}$.
We consider two possibilities for this bound:
$\int X(\R)d\R=LO$ and  $\int X(\R)d\R=sLL$.
The corresponding hyper-GGA functionals are
labeled hyper(LO) and hyper(sLL).
These correlation functionals are exact for one-electron systems, 
are size-consistent, and even if they are not so realistic \cite{hyper1}, 
they represent very good models to understand local hybrids, and to
compare the quality of the LO and sLL bounds.
The correlation energies of the atoms from He to Ar
as well as of jellium spheres (with 2, 8, 18, 20,
34, 40, 58, 92 , and 106 electrons and various $r_s$),
computed with these functionals, are reported in Table \ref{tab3}.
\begin{table}
\begin{center}
\caption{\label{tab3} Mean absolute relative errors (\%) for correlation energies of atoms from He to 
Ar  and of magic jellium spheres (js) (average for spheres of 2, 8, 18, 20, 34, 40, 58, 92 and 106 atoms) 
with different values of $r_s$. The reference data for jellium spheres are taken from Ref. 
\onlinecite{perdewjell}. Accurate atomic correlation energies are taken from Ref. \onlinecite{davidson1}.
}
\begin{ruledtabular}
\begin{tabular}{lrrrr}
 & PBE & TPSS & hyper(LO) & hyper(sLL) \\ 
\hline
atoms & 6.3 & 4.9 & 5.5 & 5.4 \\
js $r_s=1$ & 6.8 &  6.0 & 5.6 & 5.5 \\
js $r_s=2$ & 7.9 & 6.6 & 6.1 & 5.9 \\
js $r_s=3.25$ & 7.4 & 5.9 & 5.4 & 5.2 \\
js $r_s=4$ & 7.1 & 5.4 & 5.0 & 4.8 \\
js $r_s=5.62$ & 7.8 & 5.7 & 5.4 & 5.3 \\
Average js & 7.4 & 5.9 & 5.5 & 5.3 \\
\end{tabular}
\end{ruledtabular}
\end{center}
\end{table}
%
The hyper(sLL) functional systematically improves over 
hyper(LO), showing that the sLL bound may be an 
interesting tool in the construction of non-local 
functionals.
\par
In conclusion, we showed a new simple gradient-dependent bound (sLL), 
which has the same quality as the Lieb-Oxford (LO) bound 
for systems where the density varies rapidly but is significantly 
tighter for slowly-varying density cases. It is also the exact limit
for slowly-varying strongly correlated systems.
To indicate the utility of this new bound we applied it 
to both semilocal and non-local DFT, 
showing how it can be employed to construct improved exchange and
correlation functionals.
In particular, we found that the sLL bound improves the accuracy of the PBEsol functional
for both solid-state and molecules properties.
In addition, we showed that the sLL bound can be fruitfully
employed in the construction of hyper-GGAs. 
Even if the improvements of these functionals are in general small 
they are rather systematic. This proves the quality of the sLL bound
and suggests its importance for development of new functional forms.

\textbf{Computational details.}
Equilibrium lattice constants, bulk moduli, and cohesive energies 
have been calculated for 24 solids, including
Al, Ca, K, Li, Na (simple metals); Ag, Cu, Pd, Rh, V (transition metals);
LiCl, LiF, MgO, NaCl, NaF (ionic solids); AlN, BN, BP, C (insulators);
GaAs, GaP, GaN, Si, SiC (semiconductors). Reference data to construct this
set were taken from Ref. \onlinecite{sol1}. 
These calculations have been performed with the
VASP program \cite{kresse96} using PBE-PAW pseudopotentials. 
All Brillouin zone integrations were performed
on $\Gamma$-centered symmetry-reduced Monkhorst-Pack $k$-point meshes, 
using the tetrahedron method with Bl\"och corrections. 
For all the calculations a $24\times24\times24$ $k$-mesh 
grid was applied and the plane-wave cutoff was chosen to be
30\% larger than maximum cutoff defined for the pseudopotential 
of each considered atom.

Calculations for Table II have been performed
with the TURBOMOLE program package \cite{turbomole}
using a def2-TZVPP basis set \cite{basis1,basis2}.
For details about the molecular test sets, see Refs. \onlinecite{tests}.

Hyper-GGA calculations are non-self-consistent, using accurate exact exchange orbitals and
densities.

\textbf{Acknowledgments.} We thank Prof. Andreas Savin for useful discussions.

\end{document}